\documentclass[aps,prl,showpacs,superscriptaddress,twocolumn]{revtex4}
\usepackage{amsmath, amsthm, amssymb}
\usepackage{graphicx}
\usepackage{dcolumn}
\usepackage{bm}
\usepackage{xcolor}
\usepackage{cancel}
\usepackage[colorlinks, linkcolor=blue, citecolor=blue, urlcolor=blue]{hyperref}

\begin{document}

\title{Autonomous Quantum Heat Engine Enabled by Molecular Optomechanics and Hysteresis Switching}

\author{Baiqiang Zhu}
\affiliation{Quantum Institute for Light and Atoms, State Key Laboratory of Precision Spectroscopy, School of Physics and Electronic Science, East China Normal University, Shanghai 200241, China}

\author{Pierre Meystre}
\affiliation{Department of Physics and College of Optical Sciences, University of Arizona, Tucson, AZ 85721, USA.}

\author{Weiping Zhang}
\email{wpz@sjtu.edu.cn}
\affiliation{School of Physics and Astronomy, and Tsung-Dao Lee Institute, Shanghai Jiao Tong University, Shanghai 200240, China}
\affiliation{Shanghai Branch, Hefei National Laboratory, Shanghai 201315, China}
\affiliation{Shanghai Research Center for Quantum Sciences, Shanghai 201315, China}
\affiliation{Collaborative Innovation Center of Extreme Optics, Shanxi University, Taiyuan, Shanxi 030006, China}

\author{Keye Zhang}
\email{kyzhang@phy.ecnu.edu.cn}
\affiliation{Quantum Institute for Light and Atoms, State Key Laboratory of Precision Spectroscopy, School of Physics and Electronic Science, East China Normal University, Shanghai 200241, China}
\affiliation{Shanghai Branch, Hefei National Laboratory, Shanghai 201315, China}

\begin{abstract} 
By integrating molecular optomechanics with molecular switches, we propose a scheme for a molecular quantum heat engine that operates autonomously through hysteretic feedback without external driving or modulation. Through a comparative analysis conducted within both semiclassical and fully quantum frameworks, we reveal the influence of quantum properties embedded within the autonomous control elements on the operational efficiency and performance of this advanced molecular machine.
\end{abstract}

\maketitle

\emph{Introduction.}---Artificial Molecular Machines (AMMs), are bottom-up designed nanomachines aimed at mimicking the behavior of molecular proteins in living systems, efficiently converting various forms of energy into mechanical work to drive essential biological processes \cite{julicher1997modeling,schliwa2003molecular,kassem2017artificial,iino2020introduction,vale2003molecular,hernandez2004reversible,balzani2006autonomous,hugel2002single,harris2018new,torres2024bistable,avellini2012photoinduced,yao2020bistable,chen2016redox,merino2018single,goodwin2017molecular,hicks2011new,shepherd2013molecular,nguyen2007design,kornilovitch2002bistable,emberly2003smallest,martin2006mastering,thijssen2006vibrationally,litman2020multidimensional,wang2023electrical,venkataramani2011magnetic}.
Advances in nanofabrication have markedly improved our ability to detect and manipulate molecules on surfaces, facilitating the transition of AMMs from traditional solution-phase chemistry to surface and interface physics. Electric, magnetic, and optical energy have emerged as viable alternatives to chemical energy for powering AMMs. In particular, the strong coupling between molecules and light achievable in minute plasmonic cavities has led to the emerging field of molecular optomechanics \cite{benz2016single,esteban2022molecular,roelli2016molecular,ashrafi2019optomechanical,lombardi2018pulsed,schmidt2016quantum}, where optomechanical coupling strengths can exceed those of state-of-the-art microfabricated devices by several orders of magnitude and provide a promising platform to realize quantum AMMs operable at room temperature.

Power and control are two essential functional components of AMMs, resulting in considerable research interest in molecular motors \cite{julicher1997modeling,schliwa2003molecular,kassem2017artificial,iino2020introduction,vale2003molecular,hernandez2004reversible,balzani2006autonomous} and switches \cite{hugel2002single,harris2018new,torres2024bistable,avellini2012photoinduced,yao2020bistable,chen2016redox,merino2018single,goodwin2017molecular,hicks2011new,shepherd2013molecular,nguyen2007design,kornilovitch2002bistable,emberly2003smallest,martin2006mastering,thijssen2006vibrationally,litman2020multidimensional,wang2023electrical,venkataramani2011magnetic}. The former function essentially as heat engines, akin to proteins that extract energy from the environment, convert a fraction into productive motion, and dissipate the rest as heat. The latter are molecules capable of reversibly transitioning between two or more stable states in response to environmental stimuli. In this work we theoretically demonstrate a molecular optomechanical platform that integrates these two components to operate an autonomous optomechanical quantum heat engine, a key building block for the development of more intricate and functional quantum AMMs powered by light.

Much work in traditional optomechanics has been devoted to quantum versions of classical reciprocating engines \cite{blickle2012realization,quan2007quantum,feldmann2003quantum,zhang2014quantum,peterson2019experimental,rossnagel2016single,serra2016mechanical}. But due to their periodic external control consuming more energy than produced, there is a growing interest in autonomous quantum heat engines, inspiring several proposals featuring quantum versions of clocks and flywheels, analogous to classical automatic control~\cite{carollo2020nonequilibrium,elouard2015reversible,serra2016mechanical,seah2018work,mari2015quantum,tonner2005autonomous,roulet2017autonomous,toyabeme2020experimental}.

We focus here on the use of feedback control as an automatic control mechanism as there are already extensive architectures that integrate bistability and hysteresis into molecular machines to create hysteretic molecular switches~\cite{torres2024bistable,yao2020bistable,chen2016redox,goodwin2017molecular,hicks2011new,shepherd2013molecular,kornilovitch2002bistable}. These switches can establish a history-dependent negative feedback control loop for the quantum heat engine. Compared to the more common measurement feedback control \cite{zhang2017quantum}, this method avoids consuming additional energy when storing and amplifying measured signals. And importantly, optomechanical heat engines controlled by such a hysteretic molecular switch also offer a unique opportunity to investigate changes in the nature of automatic control and work extraction as they transition from the classical to the quantum regime, a consequence of their ultrasmall size and remarkable robustness to thermal noise. As we will show, when both quantum tunneling in the molecular switch and the quantum correlations between the engine and feedback controller are taken into account, the autonomous molecular engine exhibits distinct characteristics in terms of output power, operational parameter range, and plasmon statistics.

\begin{figure}
	\includegraphics[width=0.47\textwidth]{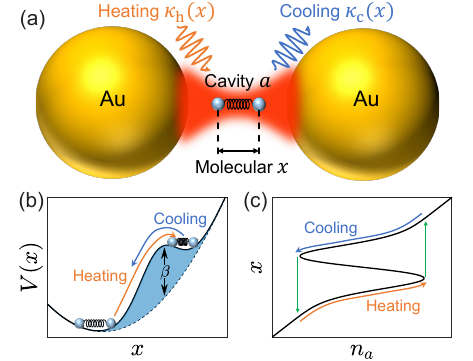}
	\centering
	\caption{(Color online) (a) Autonomous molecular heat engine with a molecular switch in a plasmonic cavity (gold spheres), coupled via dispersive and dissipative optomechanics. (b) The effective potential of the hysteretic molecular switch, where radiation pressure triggers the transition between states. (c) Heat engine's working cycle, featuring a hysteresis loop in the molecular reaction coordinate $x$, driven by radiation pressure 
 proportional to plasmon number $n_{\rm a}$. The parameters are $\theta=\hbar\omega_{\rm m}x_0^{-1}, L=x_0, \beta=2\hbar\omega_{\rm m}$, where $x_0=\sqrt{\hbar/2m\omega_{\rm m}}$ represents the vacuum fluctuation of the reaction coordinate.}
	\label{fig sketch}
\end{figure}

\emph{Model}---Hysteretic molecular switches are unique molecules that transit between stable configurations at different thresholds, depending on the direction of the switch. This results in hysteretic changes in properties such as adsorption/desorption \cite{chen2017molecular,merino2018single}, conductivity \cite{lotze2012driving,thijssen2006vibrationally,halbritter2008huge,trouwborst2009bistable,kornilovitch2002bistable,emberly2003smallest,martin2006mastering,xia2024mechanisms}, and magnetism \cite{goodwin2017molecular,sano1997molecular,torres2024bistable,hicks2011new,shepherd2013molecular,venkataramani2011magnetic} in response to chemical, electrical, or optical stimuli. The switching dynamics can be modeled by an oscillator in a double-well potential
\begin{equation}
    V(\hat{x}) = \frac{1}{2} m \omega_{\rm m}^2 \hat{x}^2 + \theta \hat{x} + \beta e^{-\hat{x}^2 / L^2},
    \label{eq bistable potential}
\end{equation}
where $\hat{x}$ is the molecular reaction coordinate, an abstract 1D representation of molecular configuration changes in response to stimuli, often related to geometric parameters such as molecular bond length or bond angle, and $m$ and $\omega_{\rm m}$ represent the oscillator's effective mass and frequency, respectively. The parameter $\theta$ accounts for the potential asymmetry, while $\beta$ and $L$ characterize the energy barrier's height and width. The two energy minima correspond to bistable molecular configurations that originate directly from distinct internal vibrational states \cite{thijssen2006vibrationally,shepherd2013molecular}, alternative molecular isomeric states \cite{hugel2002single,avellini2012photoinduced,chen2016redox,goodwin2017molecular,hicks2011new,nguyen2007design,kornilovitch2002bistable,venkataramani2011magnetic}, or varying adsorption states on surfaces \cite{merino2018single,emberly2003smallest,martin2006mastering,litman2020multidimensional,wang2023electrical,xia2024mechanisms}. As a result of the potential's asymmetry, the required stimulus energy differs during state switching, as illustrated in Fig.~\ref{fig sketch}(b).

We proceed by embedding the molecular switch within a plasmonic nanocavity crafted from gold nanospheres (Fig.~\ref{fig sketch}(a)) -- or alternative structures exhibiting surface-enhanced Raman scattering effects \cite{esteban2022molecular,roelli2016molecular,ashrafi2019optomechanical,lombardi2018pulsed,schmidt2016quantum,benz2016single} -- in such a way that the reaction coordinate aligns with the Raman-excitable molecular vibration. This results in the cavity optomechanical coupling between the local plasmon field ($\hat{E}\sim \hat{a}+\hat{a}^\dagger$) and the induced Raman dipole ($\hat d\sim \hat{x}\hat{E}$) of the molecule, and variations in the intensity of the plasmon field can then stimulate a switch of the molecule between two states of the bistable potential $V(\hat x)$. At the same time, this coupling also results in a shift and broadening in the cavity mode frequency, phenomena that are respectively referred to as dispersive and dissipative cavity optomechanical effects \cite{primo2020quasinormal,sankey2010strong,kyriienko2014optomechanics,gu2013squeezing,li2009reactive}.

When coupling this system to both a hot and a cold reservoir that exhibit distinct, narrowly peaked frequency spectra, we can then exploit the molecular switch as a feedback controller to automate the thermodynamic cycle of the optomechanical heat engine. To see how this works, assume that the cavity plasmon mode is initially at the peak frequency of the hot reservoir.  The cavity mode will therefore be energized, and the increase in radiation pressure triggers a molecular state transition via dispersive optomechanical coupling. As a result the cavity mode frequency will be shifted in such a way that it is now at the peak frequency of the cold reservoir, leading to the cooling of the cavity mode. The reduced radiation pressure triggers a reversed molecular transition back to its initial state. At that point the mode frequency has returned at the peak frequency of the hot reservoir, and the next heating-cooling sequence can proceed, establishing a feedback control loop. The hysteresis of the molecular switch in Fig.~\ref{fig sketch}(c), enabling periodic self-sustaining oscillations under suitable conditions, is key to maintaining the loop.

This heat engine can be modeled as a coupled system of three oscillators, the first one describing the molecule and the other two a pair of cavity modes of distinct frequencies, which can be realized through meticulously designed dielectric structures such as hybrid cavity-antenna resonators \cite{shlesinger2021integrated,shlesinger2023hybrid,abutalebi2024single}.
To simplify the thermodynamic analysis, we combine the two coupled cavity modes into a normal mode $\hat a$, reducing the analysis to a two-oscillator model. The full model and detailed derivation are provided in the supplementary material (SM). The dynamics of the system is then governed by the master equation
\begin{equation}
	\dot{\rho} = -\frac{i}{\hbar}[\hat{H}_{\rm om},\rho] + (\hat{\mathcal{L}}_{\rm h} + \hat{\mathcal{L}}_{\rm c} + \hat{\mathcal{L}}_{\gamma})\rho,
	\label{eq original master}
\end{equation}
where the total Hamiltonian is given by
\begin{equation}
	\hat{H}_{\rm om} = \hbar(\omega_{\rm a} + g_\omega\hat{x})\hat{a}^{\dagger}\hat{a} + \frac{\hat{p}^{2}}{2m} + V(\hat{x}),
	\label{eq total hamiltonian}
\end{equation}
where $g_\omega$ represents the dispersive optomechanical coupling strength. The Lindblad superoperators $\hat{\mathcal{L}}_{j=\rm h,c}$ describe the dissipation of the normal cavity mode $\hat a$ due to the hot and cold reservoirs, respectively. Accounting for the fact that the molecular motion alters the proportion of the two cavity modes in the normal mode $a$ results in an $x$-dependent dissipation rate—a dissipative optomechanical coupling effect— they take the form
\begin{equation}
		\hat{\mathcal{L}}_{j}\rho = (\bar{n}_{j} + 1)\mathcal{D}\Big[\sqrt{\kappa_j(\hat x)}\hat{a}\Big]\rho + \bar{n}_{j}\mathcal{D}\Big[\sqrt{\kappa_j(\hat x)}\hat{a}^{\dagger}\Big]\rho,
		\label{eq damping superoperator}
\end{equation}
$j = \{{\rm h, c}\}$, where the Lindblad dissipator is defined as $\mathcal{D}[\hat{O}]\rho = \hat{O}\rho\hat{O}^\dagger - \frac{1}{2}\{\hat{O}^\dagger\hat{O},\rho\}$, and $\bar{n}_{\rm h,c}$ are the thermal occupation numbers of the two reservoirs. As discussed in the SM, we approximate their complex expressions following the normal mode transformation by the smooth logistic functions
\begin{equation}
	\kappa_{\rm h,c}(\hat{x}) = \frac{\kappa_0}{1 + e^{\pm g_\kappa\hat{x}}}.
	\label{eq damping_rate}
\end{equation}
Here $g_\kappa$ and $\kappa_0$ denote the dissipative optomechanical coupling strength and maximum dissipation rate achievable by the cavity mode. In addition the environmental impact on the molecular switch is characterized by Brownian noise in the form of a Caldeira-Leggett superoperator,
\begin{equation}
	\hat{\mathcal{L}}_{\gamma}[\rho] = -\frac{i\gamma}{2\hbar}\Big[\hat{x},\{\hat{p},\rho\}\Big] - \frac{\gamma}{\hbar}\left(n_{\mathrm{th}} + \frac{1}{2}\right)\Big[\hat{x},[\hat{x},\rho]\Big].
	\label{eq brownian noise}
\end{equation}

In the present scheme, the cavity mode, coupled to the molecular switch and two reservoirs, serves as the working fluid of the heat engine. According to the first law of thermodynamics \cite{alicki1979quantum,kosloff2014quantum,kosloff2013quantum}, the heat engine's instantaneous power $\mathcal{P}$ is the net heat transfer rate, given by
\begin{equation}
   \mathcal{P} = \dot{Q}_{\rm h} + \dot{Q}_{\rm c}={\rm Tr}[\hat{H}_{\rm a}^{\rm eff} \hat{\mathcal{L}}_{\rm h}\rho]+{\rm Tr}[\hat{H}_{\rm a}^{\rm eff} \hat{\mathcal{L}}_{\rm c}\rho],
   \label{eq quantum power}
\end{equation}
where $\hat{H}_{\rm a}^{\rm eff} = \hbar(\omega_{\rm a} + g_\omega\hat{x})\hat{a}^\dagger\hat{a}$ is the Hamiltonian of the cavity mode. The detailed thermodynamics analysis is included in the SM. Note that there is no explicit workload or battery in this system, and all the work output is absorbed by the molecular switch, i.e., the feedback controller.

\emph{Classical switch}---For large molecules and/or high operating temperatures, the molecular switch behaves as a classical bistable oscillator. In this limit we substitute the operator $\hat{x}$ with its mean value $x$, whose evolution is governed by a classical Langevin equation which can be simplified to the overdamped equation of motion 
\begin{equation}
\dot{x} = -\frac{1}{m\gamma}(\partial_x V+\hbar g_\omega n_{\rm a}).
\label{eq position}
\end{equation}
 The evolution of the mean plasmon number, $n_{\rm a} = {\rm Tr}[\rho_{\rm a} \hat{a}^\dagger \hat{a} ]$, is then captured by the equation
\begin{equation}
\dot{n}_{\rm a} = -\kappa_0 (n_{\rm a} - \bar{n}_{\rm eff}(x) ),
\label{eq photon_num}
\end{equation}
where the equilibrium photon number $\bar{n}_{\rm eff}(x) = (\kappa_{\rm h}(x)\bar{n}_{\rm h} + \kappa_{\rm c}(x)\bar{n}_{\rm c})/\kappa_0$, depends on the reaction coordinate $x$ of the molecular switch. These coupled differential equations regulate the classical dynamics of the autonomous molecular optomechanical heat engine, with periodic, self-sustaining oscillations arising spontaneously under favorable parametric conditions, see SM for more details.

\begin{figure}
	\includegraphics[width=0.47\textwidth]{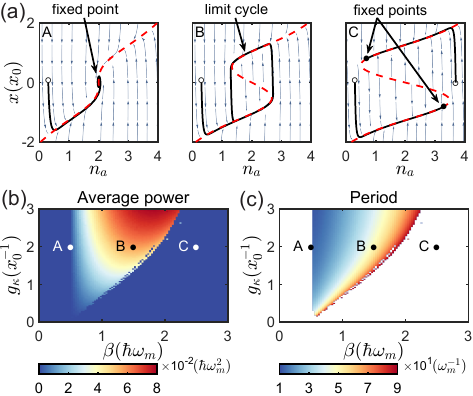}
	\centering
	\caption{(Color online) (a) Phase trajectories (black lines) in $(n_{\rm a},x)$ phase space for cases A, B, C. The hollow/solid dots mark initial/final states and the blue arrows are streamlines $(\dot{n}_{\rm a},\dot{x})$. The contour plots show the engine power $P$ (b) and working cycle period $T$ (c) as functions of $g_\kappa$ and $\beta$. The system parameters are $\kappa_{0}=0.05\omega_{\rm m}$, $\gamma=0.5\omega_{\rm m}$, $n_{\rm c}=0$, $n_{\rm h}=4$, $g_\omega=-\omega_{\rm m}x_0^{-1}$, and $\omega_{\rm a}=100\omega_{\rm m}$.}
	\label{fig classical}
\end{figure}

For a classical switch, combining Eqs.~(\ref{eq quantum power}) and (\ref{eq original master}) simplifies the expression for the power to $\mathcal{P} = -\hbar g_\omega n_{\rm a}\dot{x}$, so that the average power output of the heat engine becomes $P = \frac{1}{T} \oint \mathcal{P} dt$, where the integration encompasses a complete working cycle of the heat engine, that is one period $T$ of the self-sustaining oscillations. 

Figs.~\ref{fig classical}(b) and (c) show how the energy barrier height $\beta$ and dissipative coupling strength $g_\kappa$ affect $P$ and $T$.  Optimal engine performance requires a high $g_\kappa$ and moderate $\beta$. Low barrier heights $\beta$ weaken the hysteresis loop, while excessive $\beta$ hinders optomechanical switching. Fig.~\ref{fig classical}(a) illustrates the system dynamics at points A (low $\beta$), B (moderate $\beta$), and C (high $\beta$) in the 2D phase space $(n_{\rm a}, x)$. For B, the equilibrium positions derived from Eq.~(\ref{eq position}) form a hysteresis loop with two stable branches (red dashed line), as confirmed by the streamlines $(\dot{n}_{\rm a},\dot{x})$ (blue arrows). Any phase trajectory (black line) evolves to a limit cycle, indicating self-oscillations between branches. In contrast, for the parameters of A the system converges to a fixed point (rest state). Finally, for  C the system exhibits two fixed points, the final state depending on the specific initial conditions.

A distinct boundary exists between the operational and non-operational domains of the autonomous heat engine. This boundary can be derived through an analysis of the self-sustaining oscillation's trigger threshold from Eqs.~(\ref{eq photon_num}) and (\ref{eq position}), resulting in two specific curves, $\hbar g_\omega\bar{n}_{\rm eff}(x_\pm)+\partial_{x}V(x_\pm)=0$, where $x_\pm$ represent the two inflection points of potential $V(x)$ (see SM for more details).
The output work $W$ approaches its maximum value near this boundary, but the power $P$ decreases due to the simultaneously maximized period $T$.

\begin{figure}
	\includegraphics[width=0.47\textwidth]{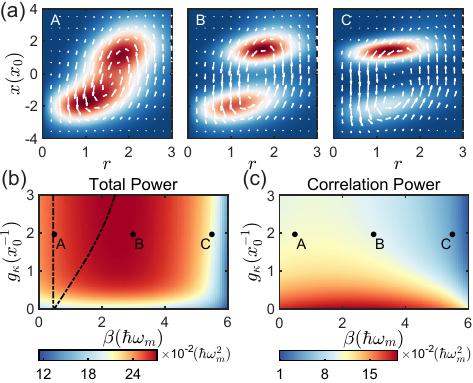}
	\centering
	\caption{(Color online) (a) Quasi-probability density $\tilde{\mathcal{Q}}(r,x)$ for cases corresponding to points A, B, and C in (b), respectively, with white arrows indicating the quasi-probability flows. The contour plots illustrate (b) the steady-state engine power and (c) the contribution from quantum correlation, within the framework of a quantum switch, where the dashed line outlines the boundary separating the classical operational domain. The parameters utilized are identical to those in Fig.~\ref{fig classical}, and $n_{\rm th}=0$.}
\label{fig quantum1}
\end{figure}

\emph{Quantum switch}---For small molecules operating at low temperatures the dynamics of the engine are governed by the full master equation~(\ref{eq original master}), which results in its evolution toward a steady state density operator $\rho_{\rm s}$. This means in particular that the mean field intensity of the plasmonic field and molecular reaction coordinate no longer exhibit a bistable behavior~\cite{drummond1980quantum}, so the heat engine can no longer be understood in terms of phase space trajectories as depicted in Fig.~\ref{fig classical}. Rather, its operation, and more specifically the exchange of heat and work that still persist under these conditions, rely now fully on the quantum properties of the molecule-field system, most importantly on quantum tunneling within the double-well potential, but also the quantum correlations that it develops with the cavity mode.

The steady-state instantaneous power $\mathcal{P}_{\rm s}$, as defined in Eq.~(\ref{eq quantum power}), is depicted in Fig.~\ref{fig quantum1}(b). In contrast to the classical case, no boundary circumscribes the operational region. This is because the molecular switch operates via quantum tunneling, even when the plasmon number $n_{\rm a}$ is insufficient to propel the molecule across the barrier $\beta$. 
Moreover, the power significantly exceeds that of the classical case due to the incorporation of both the mechanical work's power and an additional energy flow that is overlooked in the classical approach but fosters quantum correlations between the molecule and the plasmon~\cite{francica2017daemonic,manzano2018optimal}. This is exemplified by the difference $-(\hbar g_\omega/m)(\langle \hat{a}^\dagger\hat{a}\hat{p}\rangle-\langle\hat{a}^\dagger\hat{a}\rangle\langle \hat{p}\rangle)$ illustrated in Fig.~\ref{fig quantum1}(c).

Although $\mathcal{P}_{\rm s}$ is independent of time, the quantum dynamics of the engine can be described by quasi-probability distributions and flows in phase space. The former is represented by a modified Husimi $Q$-function
\begin{equation}
	\tilde{\mathcal{Q}}(r,x)=\frac{1}{\pi}\int_{0}^{2\pi}\langle{re^{i\phi},x}|\rho_{\rm s}|{re^{i\phi},x}\rangle{r}\,{\rm d}\phi,
	\label{eq phase representation}
\end{equation}
where $|re^{i\phi}\rangle$ denotes coherent states of the cavity mode with amplitude $r$ and phase $\phi$, while $|x\rangle$ represents eigenstates of the molecular reaction coordinate operator. The quasi-probability flow is characterized by the vector field $(u(r,x),v(r,x))$, where $u$ and $v$ represent flow amplitudes in the $r$ and $x$ coordinates, respectively, and are derived from (see SM for more details)
\begin{eqnarray}
u(r,x)  & =&-\int_{0}^{r}\tilde{D}(r',x){\rm d}r', \quad v(r,x)=\int_{-\infty}^{x}\tilde{D}(r,x'){\rm d}x',\nonumber \\
\tilde{D}(r,x) & =&\frac{1}{\pi}\int_{0}^{2\pi}\langle{re^{i\phi},x}|(\hat{\mathcal{L}}_{\rm h}\rho_{\rm s}+\hat{\mathcal{L}}_{\rm c}\rho_{\rm s})|{re^{i\phi},x}\rangle{r}\,{\rm d}\phi.
		\label{eq flow field}
\end{eqnarray}
Analogous to the classical case, Fig.~\ref{fig quantum1}(a) illustrates the quasi-probability distribution and flow for cases involving low, medium, and high energy barriers ($\beta$). In case A (low $\beta$), the double-well potential collapses into a single well, but non-zero displacement fluctuations sustain the quasi-probability flow, keeping the engine operational. In contrast, case C (high $\beta$) displays pronounced quasi-probability localization, as the barrier impedes probability transfer between wells, effectively stalling the engine. 
Case B (medium $\beta$) exhibits a bimodal quasi-probability distribution, where the flow lines signify the dynamics of quasi-probability exchange between the peaks, reminiscent of the hysteresis loop in the classical case.

\begin{figure}
	\includegraphics[width=0.48\textwidth]{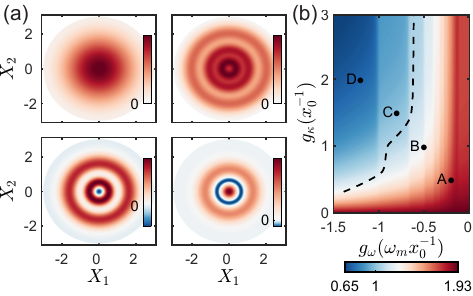}
	\centering
	\caption{(a) Wigner function of the cavity mode for four selected points indicated in (b). (b) Second-order field correlation $g^{(2)}(0)$ of the cavity mode as a function of optomechanical couplings $g_\omega$ and $g_\kappa$, with $g^{(2)}(0)=1$ indicated by a dashed line. All parameters as in Fig.~\ref{fig classical}, except for $\beta=4\hbar\omega_{\rm m}$.}
	\label{fig quantum2}
\end{figure}

In contrast to the classical scenario, where the cavity mode maintains thermal states throughout the engine's operational cycle, the quantum nature of the molecular switch results in the emergence of steady-state quantum correlations and in nonclassical statistics for the cavity mode. As illustrated in Fig.~\ref{fig quantum2}(b), the second-order correlation function, $g^{(2)}(0) = \langle\hat{a}^\dagger\hat{a}^\dagger\hat{a}\hat{a}\rangle/\langle\hat{a}^\dagger\hat{a}\rangle^2$, dips below unity as the optomechanical couplings $g_\omega$ and $g_\kappa$ intensify, a phenomenon that coincides with the presence of negative Wigner functions in Fig.~\ref{fig quantum2}(a). The negativity of the Wigner function underscores the pronounced nonclassical characteristics of the heat engine.

\emph{Conclusion.}---In summary, we have proposed and analyzed a quantum AMM that combines molecular optomechanical coupling and hysteretic molecular switches to achieve an autonomous quantum heat engine. The inherent nonclassical features of the molecular switch, encompassing quantum fluctuations, correlations, and tunneling, profoundly influence the autonomous operation and overall performance of the heat engine. This work also validates the potential of molecular optomechanical systems to serve as a platform for exploring AMMs that transcend the classical-to-quantum boundary. Future research endeavors will explore the diverse applications of molecular optomechanics in other AMMs such as molecular shuttles and molecular logic gates while also investigating their performance at the quantum scale.

\begin{acknowledgments}
\emph{Acknowledgements}---We acknowledge enlightening discussions with Lu Zhou. This work was supported by the Innovation Program for Quantum Science and Technology (No. 2021ZD0303200); the National Science Foundation of China (No. 12374328, No. 11974116, and No. 12234014); the Shanghai Municipal Science and Technology Major Project (No. 2019SHZDZX01); the National Key Research and Development Program of China (No. 2016YFA0302001); the Fundamental Research Funds for the Central Universities; the Chinese National Youth Talent Support Program, and the Shanghai Talent program.
\end{acknowledgments}

\bibliography{apssamp}

\pagebreak
\widetext
\begin{center}
	\textbf{\large Supplemental Materials}
\end{center}

\setcounter{equation}{0}
\setcounter{figure}{0}
\setcounter{table}{0}
\setcounter{page}{1}
\makeatletter
\renewcommand{\theequation}{S\arabic{equation}}
\renewcommand{\thefigure}{S\arabic{figure}}
\renewcommand{\bibnumfmt}[1]{[S#1]}
\renewcommand{\citenumfont}[1]{S#1}

\section{Model and Master Equation}
In a plasmonic nanocavity, well-designed dielectric structures (e.g., hybrid cavity-antenna architecture, see Ref[58,59] in the main text) enable strong coupling between photonic and plasmonic modes, causing spectrum splitting. A molecular switch inside alters mode frequencies, leading to resonance anticrossing and mode property shifts between photonic and plasmonic states.
The following total Hamiltonian describes this phenomenon
\begin{equation}
	\hat{H}_{\rm OM} = \hbar(\omega_A + g_A\hat{x})\hat{A}^\dagger\hat{A} + \hbar(\omega_B + g_B\hat{x})\hat{B}^\dagger\hat{B} + \frac{\hbar g}{2}(\hat{A}^\dagger\hat{B} + \hat{A}\hat{B}^\dagger) + \hat{H}_{\rm m},
\end{equation}
where $\hat{A}$ and $\hat{B}$ denote the annihilation operators for a pair of coupled plasmonic and photonic modes, respectively, with frequencies $\omega_A$ and $\omega_B$. The coefficients $g_A$, $g_B$, and $g$ represent the molecular optomechanical coupling strengths and the photon-plasmon coupling strength, respectively. Additionally, $\hat{H}_{\rm m} = \frac{\hat{p}^2}{2m} + \hat{V}(\hat{x})$ represents the Hamiltonian of the molecular switch.

By employing a Bogoliubov transformation, the Hamiltonian pertaining to the cavity modes can be diagonalized into two distinct, non-interacting bosonic normal modes,
\begin{equation}
	\hat{H}_{\rm OM} = \hbar \omega_a(\hat{x})\hat{a}^\dagger\hat{a} + \hbar\omega_b(\hat{x})\hat{b}^\dagger\hat{b} + \hat{H}_{\rm m},
\end{equation}
where the frequencies of these normal modes, $\omega_{a}$ and $\omega_{b}$, are dependent on the molecular reaction coordinate $\hat{x}$,
\begin{equation}
	\omega_{a,b} = \frac{1}{2} \left[ \omega_A + \omega_B + (g_A + g_B)\hat{x} \pm \sqrt{\left(\omega_A - \omega_B + (g_A - g_B)\hat{x}\right)^2 + g^2} \right].
\end{equation}
The Bogoliubov transformation matrix is expressed as
\begin{equation}
	\begin{bmatrix}
		\hat{a} \\ \hat{b}
	\end{bmatrix}
	=
	\begin{bmatrix}
		u_{11} & u_{12} \\ u_{21} & u_{22}
	\end{bmatrix}
	\begin{bmatrix}
		\hat{A} \\ \hat{B}
	\end{bmatrix}
	=
	\begin{bmatrix}
		\frac{-g}{\sqrt{2\Omega^2 + 2\Omega\Delta}} & \frac{\Omega + \Delta}{\sqrt{2\Omega^2 + 2\Omega\Delta}} \\
		\frac{g}{\sqrt{2\Omega^2 - 2\Omega\Delta}} & \frac{\Omega - \Delta}{\sqrt{2\Omega^2 - 2\Omega\Delta}}
	\end{bmatrix}
	\begin{bmatrix}
		\hat{A} \\ \hat{B}
	\end{bmatrix},
\end{equation}
where the matrix elements are also functions of $\hat{x}$ through the relations $\Omega = \sqrt{\Delta^2 + g^2}$ and $\Delta = \omega_A - \omega_B + (g_A - g_B)\hat{x}$. The inverse transformation is given by
\begin{equation}
	\begin{bmatrix}
		\hat{A} \\ \hat{B}
	\end{bmatrix}
	=
	\begin{bmatrix}
		u_{11} & u_{21} \\ u_{12} & u_{22}
	\end{bmatrix}
	\begin{bmatrix}
		\hat{a} \\ \hat{b}
	\end{bmatrix}.
\end{equation}

\begin{figure}[!htp]
	\centering
	\includegraphics[width=0.6\textwidth]{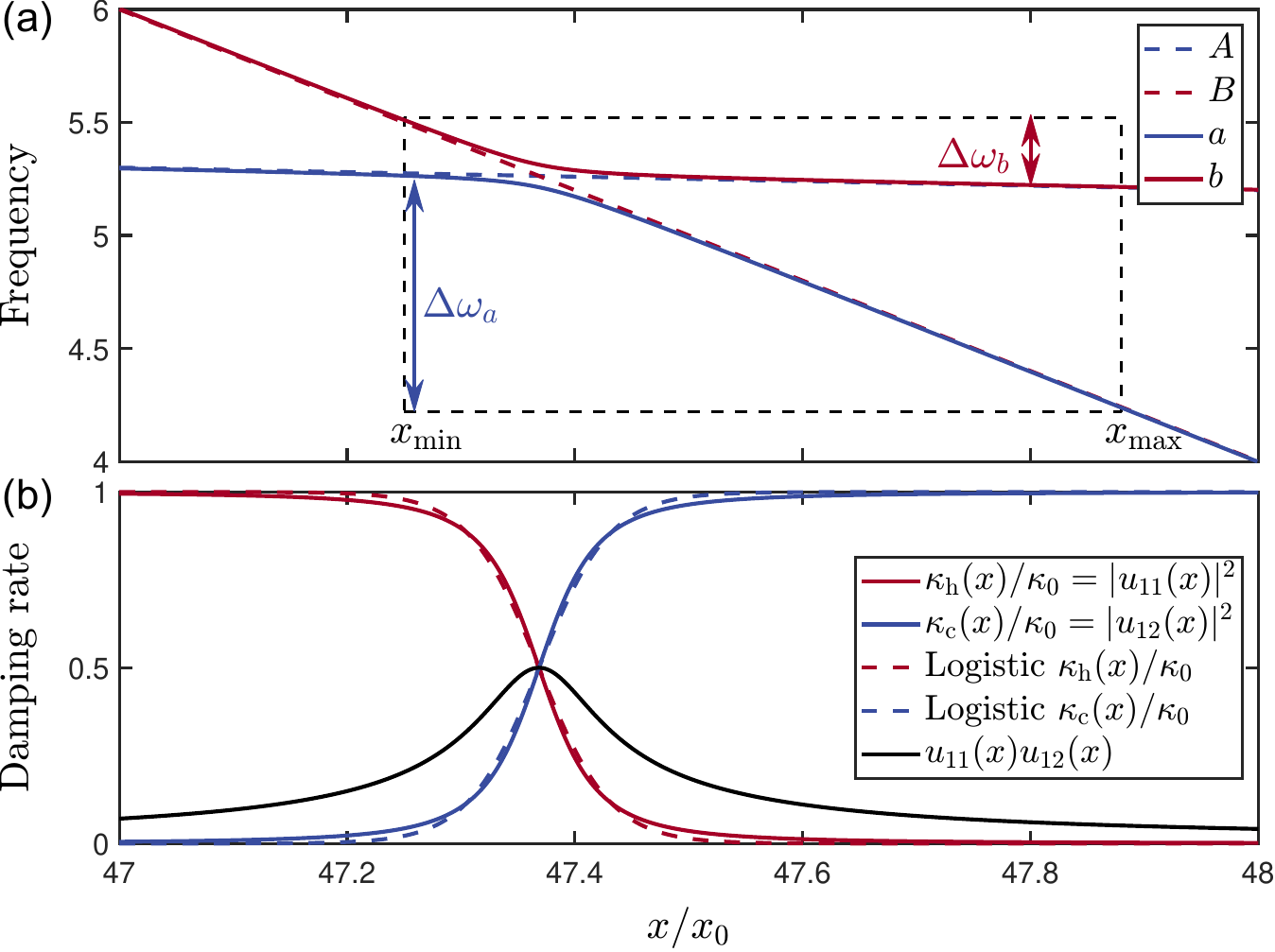}
	\caption{(color online) (a) Frequency comparison of bared modes $A, B$ (dashed lines) and normal modes $a, b$ (solid lines) as functions of the molecular reaction coordinate $x$. 
		(b) Comparison of exact (solid lines) and approximated (dashed lines) dissipation rates. The parameters are defined as $\omega_A = 10^2g$, $\omega_B = 10^3g$, $g_A = -g x_0^{-1}$, and $g_B = -20g x_0^{-1}$, where $x_0=\sqrt{\hbar/2m\omega_m}$ represents the vacuum fluctuation of the reaction coordinate.}
	\label{fig s4}
\end{figure}

As depicted in Fig. \ref{fig s4}(a), the frequencies of the normal modes $a$ and $b$ undergo transitions between photon-like and plasmon-like characteristics as $\hat{x}$ varies. Similarly, their environment transitions between one governed by a hot plasmon reservoir and another dominated by a cold photon reservoir. This behavior becomes apparent when applying Bogoliubov transformations to the dissipation dynamics of modes $A$ and $B$.

The derivation proceeds as follows: the dissipation of modes $A$ and $B$ is characterized by the application of superoperators $\hat{\mathcal{L}}_{A,B}$ on the density matrix $\rho$,
\begin{equation}
	\hat{\mathcal{L}}_{O} \rho = \kappa_O (\bar{n}_O + 1)\mathcal{D}[\hat{O}]\rho + \kappa_O \bar{n}_O \mathcal{D}[\hat{O}^\dagger]\rho, 
\end{equation}
where $\mathcal{D}[\hat{O}]\rho = \hat{O}\rho\hat{O}^\dagger - \frac{1}{2}\{\hat{O}^\dagger\hat{O}, \rho\}$ denotes the standard Lindblad dissipator for mode $O$ ($O=A, B$), and $\kappa_{A,B}$ and $\bar{n}_{A, B}$ represent the dissipation rates and thermal occupation numbers, respectively, of the corresponding modes. By employing the Bogoliubov transformation, we can express these superoperators in terms of the normal mode operators $\hat{a}$ and $\hat{b}$:
\begin{equation}
	\begin{aligned}
		\hat{\mathcal{L}}_a \rho &= \left( \kappa_A |u_{11}|^2 (\bar{n}_A + 1) + \kappa_B |u_{12}|^2 (\bar{n}_B + 1) \right)\mathcal{D}[\hat{a}]\rho \\
		&\quad + \left( \kappa_A |u_{11}|^2 \bar{n}_A + \kappa_B |u_{12}|^2 \bar{n}_B \right)\mathcal{D}[\hat{a}^\dagger]\rho, \\
		\hat{\mathcal{L}}_b \rho &= \left( \kappa_A |u_{21}|^2 (\bar{n}_A + 1) + \kappa_B |u_{22}|^2 (\bar{n}_B + 1) \right)\mathcal{D}[\hat{a}]\rho \\
		&\quad + \left( \kappa_A |u_{21}|^2 \bar{n}_A + \kappa_B |u_{22}|^2 \bar{n}_B \right)\mathcal{D}[\hat{a}^\dagger]\rho.
	\end{aligned}
\end{equation}
Note that the expressions for $\hat{\mathcal{L}}_a \rho$ and $\hat{\mathcal{L}}_b \rho$ have been refined to correctly represent the terms in accordance with the Lindblad form. 
Moreover, cross terms like $\sqrt{\kappa_A\kappa_B}u_{11}u_{21}\left(\hat{a}\rho \hat{b}^\dagger - \frac{1}{2} \{\hat{b}^\dagger, \hat{a}\} \rho - \frac{1}{2} \rho \{\hat{b}^\dagger, \hat{a}\} \right)$ have been omitted, as the coefficient $u_{11}u_{21}$ is significant only in the transient switching region, denoted by the black line in Fig.~\ref{fig s4}(b). When the cavity's dissipation rates are much smaller than the molecule's, i.e., $\kappa_{A,B}\ll \gamma$, their contributions can be safely disregarded.

Considering the dependence of $u_{ij}$ on $\hat{x}$, we define effective optomechanical dissipation rates: $\kappa_{\rm h}(\hat{x}) = \kappa_A |u_{11}|^2$ for the hot reservoir ($\bar{n}_{\rm h} = \bar{n}_A$) and $\kappa_{\rm c}(\hat{x}) = \kappa_B |u_{12}|^2$ for the cold reservoir ($\bar{n}_{\rm c} = \bar{n}_B$). The temperature difference arises from $\omega_A < \omega_B$. Based on these, we decompose $\hat{\mathcal{L}}_a$ into heating and cooling components: $\hat{\mathcal{L}}_a \rho = \hat{\mathcal{L}}_h \rho + \hat{\mathcal{L}}_c \rho$, where
\begin{equation}
	\begin{aligned}
		\hat{\mathcal{L}}_h \rho &= (\bar{n}_{\rm h} + 1)\mathcal{D}[\sqrt{\kappa_{\rm h}(\hat{x})} \hat{a}]\rho + \bar{n}_{\rm h}\mathcal{D}[\sqrt{\kappa_{\rm h}(\hat{x})} \hat{a}^\dagger]\rho, \\
		\hat{\mathcal{L}}_c \rho &=(\bar{n}_{\rm c} + 1)\mathcal{D}[\sqrt{\kappa_{\rm c}(\hat{x})} \hat{a}]\rho + \bar{n}_{\rm c}\mathcal{D}[\sqrt{\kappa_{\rm c}(\hat{x})} \hat{a}^\dagger]\rho.
	\end{aligned}
\end{equation}
This formulation facilitates the description of the coupling between the normal mode $a$ and both the hot and cold reservoirs, governed by $\hat x$, representing a dissipative optomechanical coupling.

Assuming $\kappa_A = \kappa_B = \kappa_0$ for simplicity, we derive that $\kappa_{\rm h}(\hat{x}) + \kappa_{\rm c}(\hat{x}) = \kappa_0$ given the condition $|u_{11}|^2 + |u_{12}|^2 = 1$. This indicates that the coupling strengths to the hot and cold reservoirs exhibit opposite trends as a function of $\hat{x}$.
Given the complexity of the expressions for $\kappa_{\rm h,c}(\hat{x})$, we approximate their behavior in the main text using smooth logistic functions,
\begin{equation}
	\kappa_{\rm h,c}(\hat{x}) \approx \frac{\kappa_{0}}{1 + e^{\pm g_{\kappa} \hat{x}}},
	\label{eq damping_rate}
\end{equation}
where $g_{\kappa}$ represents the effective dissipative optomechanical coupling strength. The sign ($\pm$) indicates opposite variations with $\hat{x}$, distinguishing the coupling to the hot and cold reservoirs. This approximation aligns well with the exact values (see Fig.~\ref{fig s4}(b)). 

Besides, the frequency $\omega_a(\hat{x})$ of the normal mode $a$ exhibits a monotonic dependence on $\hat{x}$ as illustrated in Fig.~\ref{fig s4}(a). For simplicity in theoretical analysis, we linearize this dependence by approximating it as
\begin{equation}
	\omega_a(\hat{x}) \approx \omega_{a} + g_\omega \hat{x},
\end{equation}
where $g_\omega$ represents the effective dispersive optomechanical coupling strength and $\omega_{a}$ represents the base frequency of mode $\hat{a}$, respectively.

While the expressions for $\hat{\mathcal{L}}_b$ and $\omega_b(\hat{x})$ of normal mode $b$ bear a resemblance to those of normal mode $a$, their variation with respect to $x$ is opposite to that of normal mode $\hat{a}$. As depicted in Fig.~\ref{fig s4}(a), when the operational range of $x$ is asymmetric relative to the spectral anticrossing point, the frequency shift of mode $\hat{b}$ is notably smaller compared to that of mode $\hat{a}$. Therefore, we exclude mode $\hat{b}$ from our subsequent analysis and derive the master equation of the system as in the main text,
\begin{equation}
	\begin{aligned}
		\dot{\rho} = -\frac{i}{\hbar} [\hat{H}_{\rm om}, \rho] + (\hat{\mathcal{L}}_{\rm h} + \hat{\mathcal{L}}_{\rm c} + \hat{\mathcal{L}}_{\gamma})\rho,
		\label{eq master}
	\end{aligned}
\end{equation}
where $\hat{\mathcal{L}}_\gamma \rho$ describes the Brownian dissipation of the molecular switch.

\section{Classical Dynamics}
When we neglect the quantum correlation between the molecular switch and the cavity field, allowing the joint density matrix to be factorized as $\rho(t) \approx \rho_{\rm a}(t) \otimes \rho_{\rm m}(t)$, the master equation (\ref{eq master}) can be decoupled into two separate reduced master equations, one for the cavity mode and the other for the molecular switch, 
\begin{equation}
	\begin{aligned}
		\dot{\rho}_{\rm a} &\approx -i\left[(\omega_{\rm a} + g_\omega \langle\hat{x}\rangle)\hat{a}^\dagger\hat{a}, \rho_{\rm a}\right] + \langle\hat{\kappa}_{\rm h}\rangle((\bar{n}_{\rm h} + 1)\mathcal{D}[\hat{a}]\rho_{\rm a} + \bar{n}_{\rm h}\mathcal{D}[\hat{a}^\dagger]\rho_{\rm a}) + \langle\hat{\kappa}_{\rm c}\rangle ((\bar{n}_{\rm c} + 1)\mathcal{D}[\hat{a}]\rho_{\rm a} + \bar{n}_{\rm c}\mathcal{D}[\hat{a}^\dagger]\rho_{\rm a}), \\
		\dot{\rho}_{\rm m} &\approx -\frac{i}{\hbar}\left[\frac{\hat{p}^2}{2m} + \hat{V}(\hat{x}) + \hbar g_\omega \langle\hat{a}^\dagger\hat{a}\rangle\hat{x}, \rho_{\rm m}\right]  -\frac{i\gamma}{2\hbar}[\hat{x},\{\hat{p},\rho_{\rm m}\}]-\frac{\gamma}{\hbar}\left(n_{\mathrm{th}}+\frac{1}{2}\right)[\hat{x},[\hat{x},\rho_{\rm m}]]
		\label{eq coupled master}
	\end{aligned}
\end{equation}
where we ignore the small contribution terms like ${\rm Tr}_a[\mathcal{D}[\sqrt{\kappa(\hat x)}\hat a] \rho_a\rho_m]$ in $\dot\rho_{\rm m}$ due to the condition $\gamma\gg\kappa_0$.

Furthermore, if we neglect their quantum fluctuations, the dynamics of the system can be described by a set of coupled equations involving the classical mean values $n_{\rm a}$, $x$, and $p$,
\begin{equation}
	\begin{aligned}
		\dot{n}_{\rm a}&={\rm Tr}[\hat{a}^\dagger \hat{a}\dot\rho_{\rm m}]=-\kappa_{\rm h}(x)(n_{\rm a}-\bar{n}_{\rm h})-\kappa_{\rm c}(x)(n_{\rm a}-\bar{n}_{\rm c}),\\
		\dot{x}&={\rm Tr}[\hat x \dot\rho_{\rm m}]=\frac{p}{m},\\
		\dot{p}&={\rm Tr}[\hat p \dot\rho_{\rm m}]=-\partial_xV(x)-\hbar g_\omega n_{\rm a}-\gamma p. 
	\end{aligned}
\end{equation}
When the molecular damping $\gamma\gg\kappa_0,\omega_m$, we can adiabatically eliminate the equation for $p$, and obtain the equations of motion presented in the main text,
\begin{equation}
	\begin{aligned}
		\dot{n}_{\rm a}&=-\kappa_0(n_{\rm a}-\bar{n}_{\rm eff}(x)),\\
		\dot{x}&=-\frac{1}{m\gamma}(\partial_x V(x)+\hbar g_\omega n_{\rm a}),
		\label{eq classical_2}
	\end{aligned}
\end{equation}
where 
\begin{equation}
	\bar{n}_{\rm eff}(x) = [\kappa_{\rm h}(x)\bar{n}_{\rm h} + \kappa_{\rm c}(x)\bar{n}_{\rm c}]/\kappa_0.
\end{equation}

\begin{figure}[!htp]
	\includegraphics[width=1.0\textwidth]{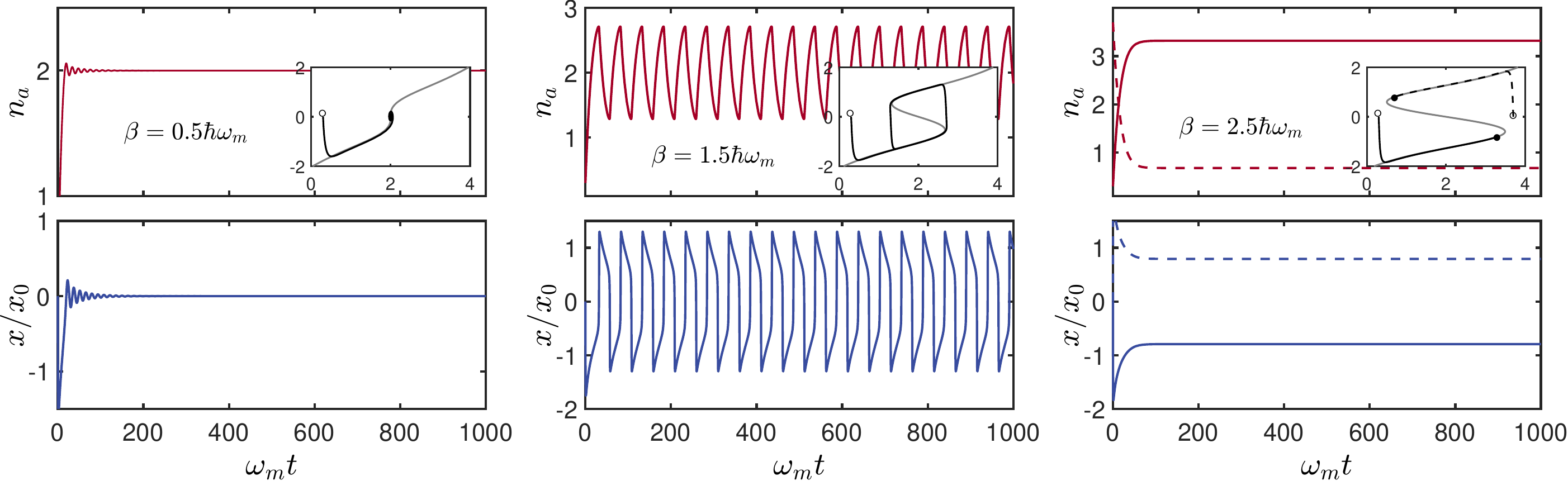}
	\centering
	\caption{Time evolution of $x$ and $n_{\text{a}}$ is shown for different potential barrier heights: $\beta = 0.5\hbar\omega_m$ (left), $\beta = 1.5\hbar\omega_m$ (middle), and $\beta = 2.5\hbar\omega_m$ (right), with the corresponding phase-space trajectories depicted in the insets. Other parameters remain the same as in Figure 2 of the main text. }
	\label{fig s1}
\end{figure}

Fig.~\ref{fig s1} illustrates the dynamical behavior of the system, governed by Eq.~(\ref{eq classical_2}), for three distinct potential barrier heights denoted as $\beta$, as detailed in the main text. 
In the absence of a potential barrier, $x$ and $n_{\text{a}}$ exhibit rapid relaxation, ultimately converging to a fixed point in the phase space $(n_{\text{a}}, x)$, as depicted in the left panel.
As the potential barrier increases, $x$ and $n_{\text{a}}$ undergo periodic oscillations, corresponding to a limit cycle within the phase space, as shown in the middle panel. This represents the system alternating between two distinct steady states through self-sustained oscillations.
When the potential barrier becomes excessively high, $x$ and $n_{\text{a}}$ once again undergo rapid relaxation but remain in one of the two fixed points, determined by the initial conditions, as illustrated in the right panel.

By conducting a stability analysis, we can delineate the boundaries of these three distinct dynamic regions within the parameter space $(g_\kappa, \beta)$, which correspond to the thresholds at which self-sustained oscillations emerge. To accomplish this, we introduce small perturbations $(\delta_n, \delta_x)$ to the steady-state solutions $(n_s, x_s)$ of Equation~(\ref{eq classical_2}). Neglecting higher-order infinitesimals, we obtain the matrix equation
\begin{equation}
	\partial_t\binom{\delta_{n}}{\delta_{x}}=\binom{-\kappa_0 \,\,\, \kappa_0\partial_x \bar{n}_{\rm eff}(x_s)}{-\frac{\hbar g_\omega}{m\gamma}\,\,\, -\frac{1}{m\gamma} \partial_x^2 V(x_s)}\binom{\delta_{n}}{\delta_{x}}.
\end{equation}
The steady solution is stable only when both eigenvalues of the coefficient matrix are non-positive. 

In cases where the molecular damping is significantly larger, specifically $\gamma \gg \kappa_0$, the eigenvalues can be approximated as $\lambda_1 \approx 0$ and $\lambda_2 \approx -\frac{1}{m\gamma} \frac{\partial^2 V(x_s)}{\partial x^2}$. Consequently, the threshold values for stable $x_s$ are determined by the equation $\frac{\partial^2 V(x_s)}{\partial x^2} = 0$, yielding two solutions: $x_s = x_\pm$. The boundary of the self-sustained oscillation region in the parameter space $(g_\kappa, \beta)$, as depicted in Fig. 2 of the main text, can be established by substituting $x_\pm$ into Eq.~(\ref{eq classical_2}), which results in 
\begin{equation}
	\hbar g_\omega \bar{n}_{\rm eff}(x_\pm) + \frac{\partial V(x_\pm)}{\partial x} = 0.
\end{equation}

\section{Thermodynamic Analysis}
In our heat engine model, the cavity mode field serves as the working medium, while the reaction coordinates of the molecular switch represent the working degree of freedom. Thus, the effective Hamiltonian of the working medium is given by
\begin{equation}
	\hat{H}_{\rm a}^{\rm eff} = \hbar(\omega_{\rm a} + g_\omega \hat{x})\hat{n}_{\rm a}.
\end{equation}
Based on the first law of thermodynamics and the definitions of work and heat within the framework of quantum thermodynamics, the instantaneous power output $\mathcal{P}$ of the heat engine is given by the sum of the heat exchange rates with the hot and cold reservoirs, 
\begin{equation}
	\mathcal{P} = \dot{Q}_{\rm h} + \dot{Q}_{\rm c} = \text{Tr}[\hat{H}_{\rm a}^{\rm eff} (\hat{\mathcal{L}}_{\rm h} + \hat{\mathcal{L}}_{\rm c}) \rho].
	\label{eq power}
\end{equation}

Using the master equation in Eq.~(\ref{eq master}), we derive:
\begin{equation}
	(\hat{\mathcal{L}}_{\rm h} + \hat{\mathcal{L}}_{\rm c}) \rho = \dot\rho + \frac{i}{\hbar}[\hat H_{\rm om}, \rho] - \hat{\mathcal{L}}_{\gamma}\rho.
\end{equation}
Substituting this into Eq.~(\ref{eq power}), and utilizing the relationship $\hat{H}_{\rm om} = \hat{H}_{\rm a}^{\rm eff} + \hat H_{\rm m}$ along with the property of the Brownian dissipation superoperator, $\text{Tr}[\hat{H}_{\rm a}^{\rm eff} \hat{\mathcal{L}}_\gamma \rho] = 0$, we obtain a revised expression for the power,
\begin{equation}
	\begin{aligned}
		\mathcal{P} &= \text{Tr}[\hat{H}_{\rm a}^{\rm eff}\dot{\rho}] + \frac{i}{\hbar} \langle [\hat{H}_{\rm a}^{\rm eff}, \hat H_{\rm om}] \rangle \\
		&= \frac{d\langle \hat{H}_{\rm a}^{\rm eff}\rangle}{dt} - \frac{\hbar g_\omega}{m} \langle \hat{n}_{\rm a} \hat{p} \rangle,
		\label{eq ins_power}
	\end{aligned}
\end{equation}
where the first term is the rate of change of internal energy, and the second term represents power done on the molecular switch. We show below that the first term does not contribute to average power and highlight differences in the second term between classical and quantum cases.

In the classical case, the operation of the heat engine is periodic, thus the average power is defined over a complete cycle,
\begin{equation}
	P = \frac{1}{T}\oint \mathcal{P} \, dt,
\end{equation}
where the integral is taken over one cycle period $T$.
Utilizing Eq.~(\ref{eq ins_power}) and disregarding quantum fluctuations and correlations, we obtain
\begin{equation}
	P = \frac{1}{T}\oint \dot{H}_{\rm a}^{\rm eff} \, dt - \frac{1}{T}\frac{\hbar g_\omega}{m} \oint n_{\rm a} p \, dt,
\end{equation}
where the first term on the right-hand side disappears due to the periodic nature of energy changes. Notably, if we substitute $p=m\dot x $ and define $F=-\hbar g_\omega n_{\rm a}$, the second term takes the form of the power exerted by the radiation pressure force, $\frac{1}{T}\oint Fdx$. 

In the quantum case, the system attains a steady state characterized by $\dot{\rho}_{\rm s} = 0$. Consequently, the term representing the internal energy change in Eq.~(\ref{eq ins_power}) diminishes. Hence, the steady-state power is expressed as
\begin{equation}
	\mathcal{P}_s = -\frac{\hbar g_\omega}{m} \langle \hat{n}_{\rm a} \hat{p} \rangle.
\end{equation}
Distinct from the classical case, the steady-state power in the quantum realm is influenced by the correlation between the cavity mode and the molecular switch. This component of the power, termed ``correlation power'' in the main text, can be measured by the difference
\begin{equation}
	\Delta \mathcal{P}_s=-\frac{\hbar g_\omega}{m} (\langle \hat{n}_{\rm a} \hat{p} \rangle - \langle \hat{n}_{\rm a} \rangle \langle \hat{p} \rangle).
\end{equation}

\section{Quasi-probability Picture}
In the quantum case, the heat engine operates in a steady state, with continuous energy exchange between the cavity field and the molecule to maintain equilibrium. We can visualize these dynamics using quasi-probability distribution and flow fields in phase space.

The full phase space of the cavity-molecule coupled system is four-dimensional. However, a reduced phase space focusing on the amplitude of the cavity field and the molecule's reaction coordinate is sufficient to capture the main quantum dynamics and enable comparisons with the classical case. Thus, we represent the quasi-probability distribution of the steady state $\rho_s$ using a modified Husimi-Q function based on the coherent state $ |\alpha = r e^{i\phi}\rangle $ with a phase integral and coordinate eigenstate $|x\rangle$, defined as
\begin{equation}
	\tilde{\mathcal{Q}}(r, x) = \frac{1}{\pi} \int_{0}^{2\pi} \langle \alpha | \langle x | \rho_{\rm s} | x \rangle| \alpha \rangle  r d\phi.
	\label{eq visua}
\end{equation}

In the case of overdamping, where $\gamma \gg \kappa_0$, the molecular switch adiabatically tracks the changes in the cavity mode and settles into the ground state $|g(n)\rangle$ of the effective potential $V(\hat{x}) + \hbar g_\omega n\hat{x}$. Subsequently, the steady state of the system can be approximately expressed as
\begin{equation}
	\rho_s \approx \sum_n P_n |n\rangle\langle n| \otimes |g(n)\rangle\langle g(n)|,
\end{equation}
where $|n\rangle$ denotes the number state of the cavity mode with a probability $P_n$. Given this, the term $[H_{\rm om}, \rho_s] = 0$, and the master equation for the steady state simplifies to
\begin{equation}
	(\hat{\mathcal{L}}_{\rm h} + \hat{\mathcal{L}}_{\rm c} + \hat{\mathcal{L}}_\gamma)\rho_{\rm s} = 0.
	\label{eq steady master equation}
\end{equation}

The superoperators $\hat{\mathcal{L}}_{\rm h,c}$ and $\hat{\mathcal{L}}_\gamma$ give rise to variations in the amplitude of the cavity field and the molecular displacement, respectively.
These variations manifest as two non-zero time derivatives of the quasi-probability distribution, which are defined as 
\begin{equation}
	\begin{aligned}
		\tilde{D}(r, x) &= \frac{1}{\pi} \int_{0}^{2\pi} \langle \alpha|\langle x | (\hat{\mathcal{L}}_{\rm h} \rho_{\rm s} + \hat{\mathcal{L}}_{\rm c}\rho_{\rm s}) |  x \rangle |\alpha\rangle r d\phi ,\\
		\tilde{M}(r, x) &= \frac{1}{\pi} \int_{0}^{2\pi} \langle \alpha | \langle x | \hat{\mathcal{L}}_{\rm \gamma} \rho_{\rm s} |x\rangle | \alpha \rangle r d\phi .
	\end{aligned}
	\label{eq u1}
\end{equation}
Given the steady state condition Eq.~(\ref{eq steady master equation}), it follows that $\tilde{M}(r, x) = -\tilde{D}(r, x)$. 

With these definitions, we can introduce a vector field $\vec{F} = (u, v)$ that delineates the quasi-probability flow within phase space $(r, x)$. The components of this vector field, $u(r', x')$ and $v(r', x')$, signify the flow density in the $r$ and $x$ dimensions, respectively, at any arbitrary position $(r', x')$ in phase space. These components are expressed through the integrals of $\tilde{D}(r, x)$ along the $r$-dimension and $\tilde{M}(r, x)$ along the $x$-dimension, respectively, as follows
\begin{equation}
	\begin{aligned}
		u(r', x') &= -\int_0^{r'} \tilde{D}(r, x') \, dr, \\
		v(r', x') &= -\int_{-\infty}^{x'} \tilde{M}(r', x) \, dx=\int_{-\infty}^{x'} \tilde{D}(r', x) \, dx.
		\label{eq flow field}
	\end{aligned}
\end{equation}
Note that a key feature of this quasi-probability flow field is its zero divergence,
\begin{equation}
	\begin{aligned}
		\nabla \cdot \vec{F} = \partial_r u(r, x) + \partial_x v(r, x) = 0,
		\label{eq div}
	\end{aligned}
\end{equation}
reflecting the conservation of probability, i.e., ${\rm Tr}\rho_{\rm s} = 1$.

\end{document}